\documentclass[journal]{IEEEtran}
\usepackage{graphicx}
\usepackage{subfig}
\usepackage{booktabs}
\usepackage{tabularx}
\usepackage{amsmath}
\usepackage{amssymb}

\ifCLASSINFOpdf
\else
\fi
\begin{document}
\title{Readout Electronics for CEPC Semi-Digital Hadron Calorimeter Pre-prototype}

\author{Yu Wang, Shubin Liu, Changqing Feng, Zhongtao Shen, Junbin Zhang, Daojin Hong, Jianbei Liu, Yi Zhou% <-this % stops a space
\thanks{Manuscript submitted Jun 24, 2018; Minor revised Nov 06, 2018. This study was supported by National Key Programme for S\&T Research and Development (Grant NO.: 2016YFA0400400) and National Science Natural Science Foundation of China (Grant No.11635007).}% <-this % stops a space
\thanks{Yu Wang, Shubin Liu, Changqing Feng, Zhongtao Shen, Junbin Zhang, Daojin Hong, Jianbei Liu, Yi Zhou are with State Key Laboratory of Particle Detection and Electronics, University of Science and Technology of China, No.96, Jinzhai Road, Hefei, Anhui, China (Corresponding Author Shubin Liu, e-mail: liushb@ustc.edu.cn ).}% <-this % stops a space
}


\markboth{}%
{Shell \MakeLowercase{\textit{et al.}}: Bare Demo of IEEEtran.cls for IEEE Journals}



\maketitle

\begin{abstract}
CEPC (Circular Electron and Positron Collider) is a large experiment facility proposed by Chinese particle physics community. One of its running option is being the Higgs factory. Calorimeter is the main part of this experiment to measure the jet energy. Semi-digital hadron calorimeter (SDHCAL) is one of the options for the hadron measurement. GEM detector with its high position resolution and flexible configuration is one of the candidates for the active layer of the SDHCAL. The main purpose of this paper is to provide a feasible readout method for the GEM-based semi-digital hadron calorimeter. A small-scale prototype is designed and implemented, including front-end board (FEB) and data interface board (DIF). The prototype electronics has been tested. The equivalent RMS noise of all channels is below 0.35fC. The dynamic range is up to 500fC and the gain variation is less than 1\%. The readout electronics is applied on a double-layer GEM detector with 1cm$\times$1cm readout pad. Result indicates that the electronics works well with the detector. The detection efficiency of MIP is over 95\% with 5fC threshold.
\end{abstract}


% \fi
\IEEEpeerreviewmaketitle

\section{Introduction}
\IEEEPARstart{D}{iscovery} of the Higgs boson leads to great interest in collider experiments. Except LHC, scientists are proposing to build next generation collider such as International Linear Collider (ILC), Compact Linear Collider (CLIC), Future Circular Collider (FCC) and Circular Electron and Positron Collider (CEPC). Especially for CEPC, it can operate at 240GeV as a Higgs factory~\cite{bib:bib1,bib:bib2}. Calorimeters play an important role in modern collider system. To achieve required energy resolution, a technique known as Particle Flow Algorithms (PFA) is applied on the calorimeters~\cite{bib:bib3}. This algorithm both combine the tracker information and energy information to reconstruct each particle individually. In order to assign the energy to the corresponding reconstructed particles correctly, this algorithm requires the calorimeters to have extremely fine segmentation both laterally and longitudinally and this leads to large numbers of readout electronics channels. Sampling calorimeters composed of active and absorber layers are usually used to perform the PFA. 

For Hadron measurement, it would cost a lot to readout each channel with high-resolution ADC. One appropriate and affordable choice is using gaseous detector with digital (1-bit) and semi-digital (2-bit) readout, so called the digital hadron calorimeter (DHCAL) and the semi-digital hadron calorimeter (SDHCAL) respectively~\cite{bib:bib4,bib:bib5}. The active elements of DHCAL can be Glass Resistive Plate Chamber (GRPC), MicroMegas and Gas Electron Multiplier (GEM)~\cite{bib:bib6}. The DHCAL and SDHCAL prototype based on GRPC have been designed and tested by CALICE group~\cite{bib:bib7,bib:bib8,bib:bib9}. The test results show that the digital readout and semi-digital readout are potential methods for hadron calorimeter.

As one of the candidates for DHCAL active layer, large area GEM detector is proposed in the CEPC pre-R\&D project. It can offer high position resolution and flexible configuration, which meets the requirement of high granularity readout. Due to the fast rise time and short recovery time, GEM detector can work in a higher counting rate compared to GRPC. Moreover, it can be built at low costs, which makes it possible in large area construction. Therefore, it is necessary to design the corresponding semi-digital readout electronics and verify the feasibility of the readout structure for the GEM detector.

In this study, the design and implementation of a semi-digital readout electronics for GEM detector is described. The readout anode pad for the GEM detector is $1cm\times1cm$, which meets the requirement of the PFA calorimeter. The main purpose is to test the performance of the front-end electronics as well as the detector. The noise and gain tests are accomplished for the front-end electronics and the results show that the dynamic range can cover the signal range of the detector. As for the detector, the main performance tests are about the response of the Minimum Ionization Particles (MIPs) and gas gain. These two parameters will have a great influence on the energy reconstruction of particles. This system is one of the pre-prototype research for the Digital Hadron Calorimeter for CEPC.

\section {Description of the Readout Electronics Design}
\begin{figure*}[htb]
\centering
\includegraphics[width=7.16in]{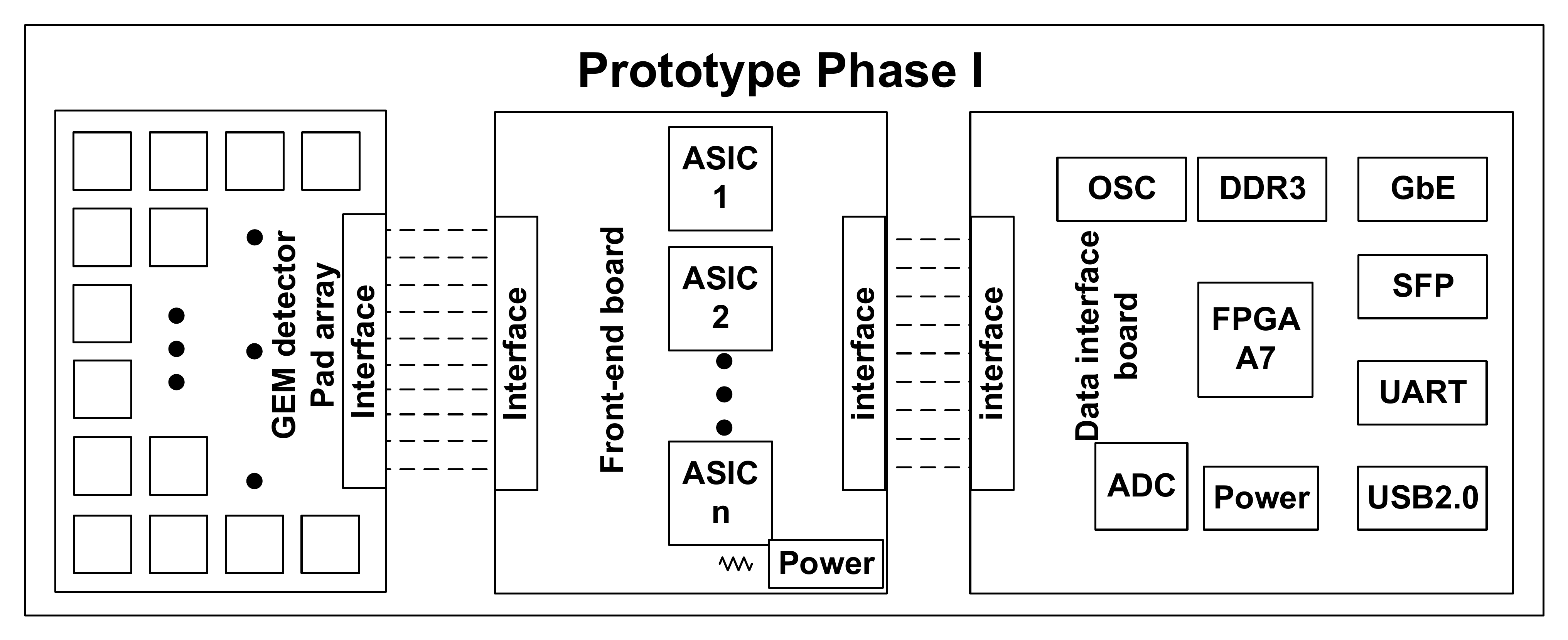}
\caption{Structure diagram of the readout electronics. The prototype contains readout plane for GEM detector, FEB and DIF.}
\label{fig:Fig1}
\end{figure*}

A small-scale prototype is designed to verify the readout electronics and test the performance of the GEM detector. This prototype contains a detector readout plane with $1cm\times1cm$ anode pads, a front-end board (FEB) with readout ASICs (Application Specific Integrated Circuit) and a data interface board (DIF) with multiple readout interfaces (Fig.~\ref{fig:Fig1}). There are several digital readout ASICs designed for the Micro Pattern Gas Detector (MicroMegas and GEM), such as GASTONE~\cite{bib:bib10}, DIRAC~\cite{bib:bib11,bib:bib12}, DCAL~\cite{bib:bib13} and MICROROC~\cite{bib:bib14,bib:bib15}.
\begin{figure*}[htb]
\centering
\includegraphics[width=7.16in]{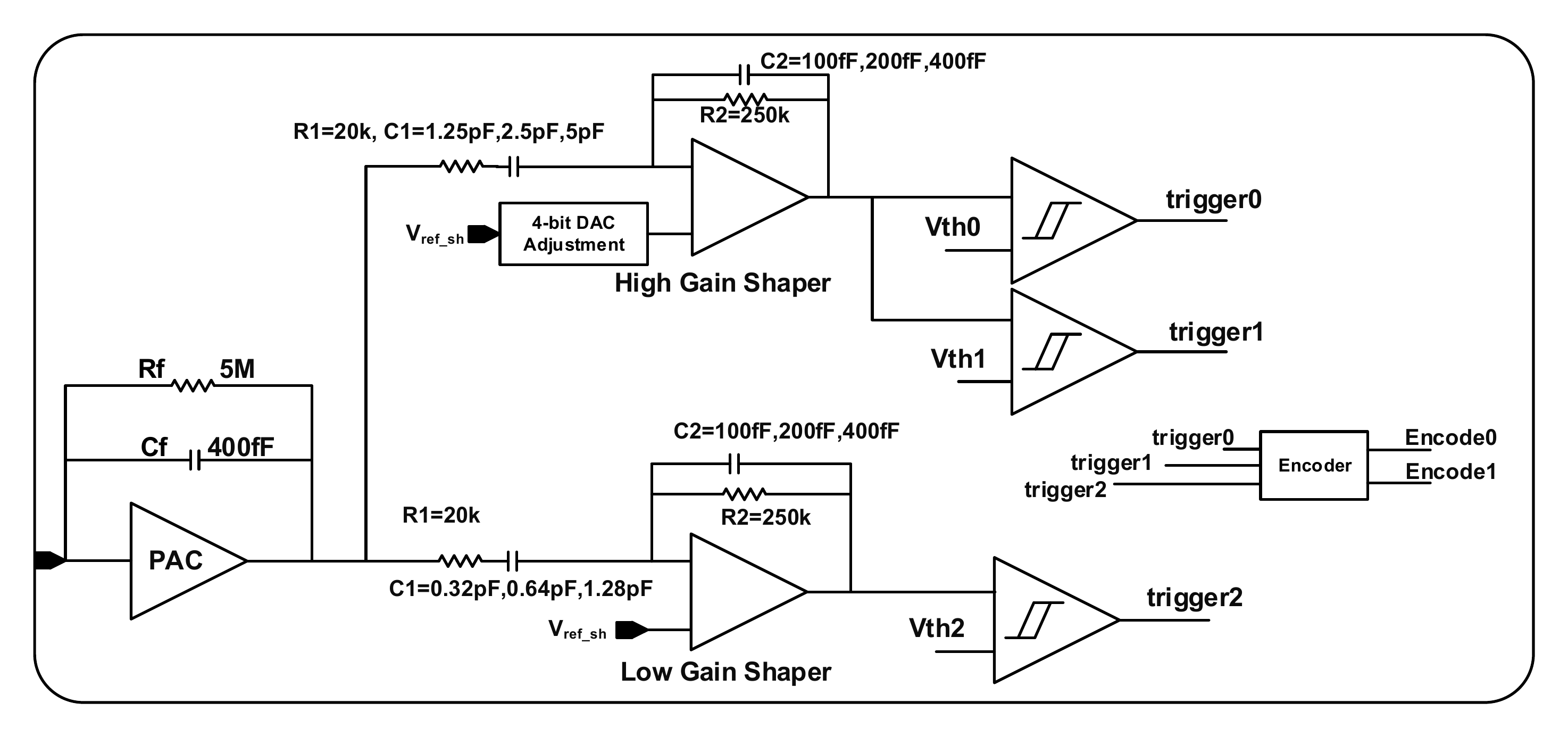}
\caption{Analog structure of MICROROC ASIC. The main parts are a low noise charge preamplifier, a high gain and a low gain CR-RC shaper and three discriminators.}
\label{fig:Fig2}
\end{figure*}
Among these ASICs, MICROROC has a dynamic range from 1fC to 500fC and is more suitable for the GEM detector readout. The specifics of the MICROROC chip and readout electronics are described below.

\subsection{Introduction to the MICROROC ASIC}
MICROROC (MICRO mesh gaseous structure Read-Out Chip) is a 64-channel readout ASIC designed at IN2P3 by OMEGA and LAPP microelectronics group~\cite{bib:bib14,bib:bib15}. The analog part of each channel in MICROROC is composed of a low noise charge-sensitive amplifier, two fast shapers of low and high gain with tunable peaking time and three discriminators for semi-digital readout (Fig.~\ref{fig:Fig2}).

The signal of the detector is collected by the charge sensitive amplifier (CSA) and then fed to the two shapers. The high gain shaper is connected to two discriminators with low and medium thresholds. The low gain shaper has a linear dynamic range up to 500fC and is connected to the discriminator with the highest threshold. The thresholds of discriminators are set by three 10-bit DACs for 64 channels. Each channel has a 4-bit DAC to shift the pedestal voltage and minimize the channel dispersions.

If one of the channels has a charge signal above the lowest threshold, the compared results of all discriminators are encoded into 2 bits individually and stored in a RAM together with the timestamp.  
\begin{figure*}[htb]
\centering
\includegraphics[width=7.16in]{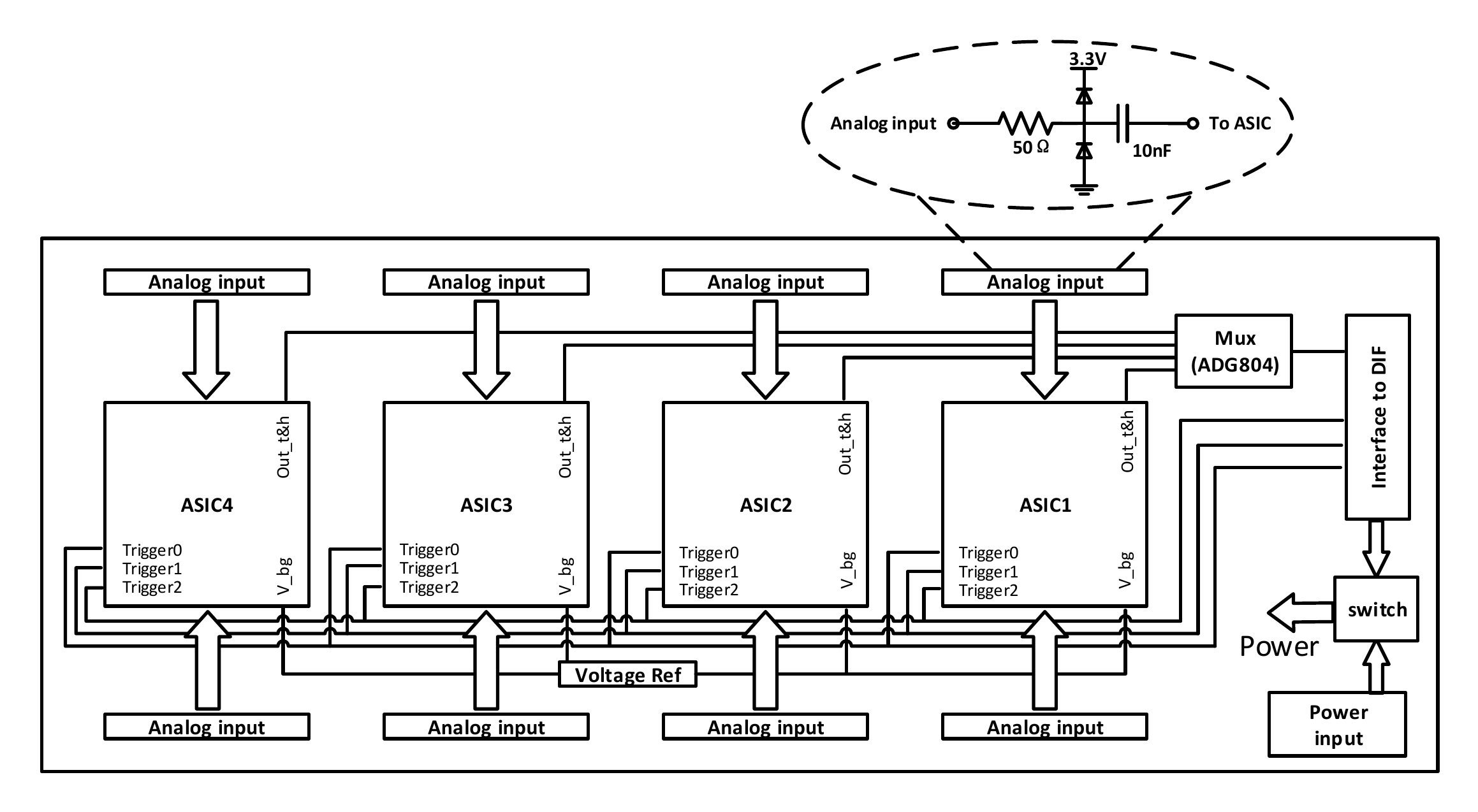}
\caption{Structure diagram of the front-end board. The charge protection circuit is shown in the top right corner.}
\label{fig:Fig3}
\end{figure*}
After acquisition, the RAM is readout by the pulse of slow clock synchronously.

\subsection{Front-end board}
As shown in Fig.~\ref{fig:Fig3}, the front-end board accommodates 4 MICROROC ASICs controlled in daisy chain. The analog signals are received from both long sides of the board. A special spark protection circuit is designed in the input stage of each channel and the analog signal is clamped between -0.7V to 4.0V by a pair of diodes. A voltage reference chip is designed to provide the external reference for bandgap voltage (if needed).

The control and test signal is connected to the data interface board through flexible plate made of kapton. There are two kinds of control signals classified by their speed: fast control and slow control. The fast control signals, which include start acquisition, trigger and reset, are sent to each MICROROC directly. While the slow control signals are in daisy chain cascade to the ASICs and configure the 592-bit registers on the chips. The slow control parameters contain all the configurations such as peaking time, thresholds setting, channel mask, etc. The digitized output data of MICROROC are sent to the DIF through the open collector (OC) gate. Benefit from the OC gate and daisy chain cascade, the ASICs on the FEB can be increased without changing the definition of the connector to DIF. Besides the control and data signals, one analog test signal and three digital signals are connected to the DIF for debugging, which are the reserved functions of MICROROC. The peak voltage of the low gain shaper can be sent to the test pin of MICROROC, which is sampled and digitized by an ADC on the DIF board. This function is useful when checking the gain of GEM detector.

The MICROROC requires a supply voltage of 3.3V, which can be obtained either from the on-board low-dropout (LDO) regulator or from the DIF.

\subsection{Data interface board}
The DIF both controls the FEB and gathers the data. It consists of three main parts: the controller part, the interface part and the power delivery part. The controller part is mainly composed of an FPGA (Field Programmable Gate Array) and the relative circuit. The function of the FPGA is to implement the control logic for FEB and communication interface. The interface part contains USB port, SFP port, DTCC (Data, Trigger, Clock and Control) port and Ethernet port. The power delivery part is implemented with the DC input port, DC-DC converters and LDO regulators. This part supports the power supply for the DIF and deliver the 3.3V power to FEB.
\begin{figure}[htb]
\centering
\includegraphics[width=3.5in]{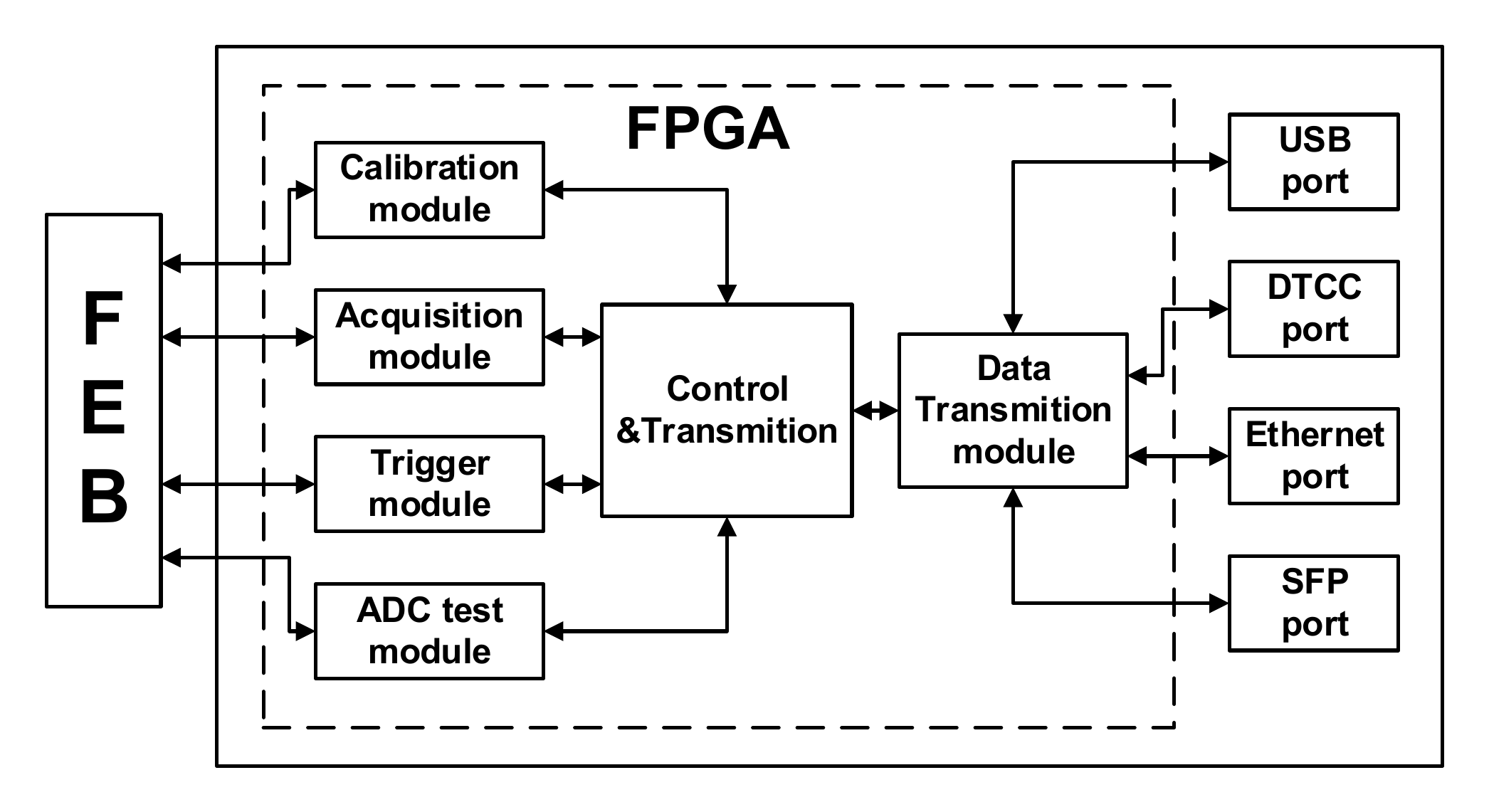}
\caption{Diagram of logic implemented in the FPGA.}
\label{fig:Fig4}
\end{figure}

The diagram of the DIF with the FPGA logic is presented in Fig.~\ref{fig:Fig4}. The calibration module is used to implement the ¡¥S-Curve¡¦ calibration for MICROROC, which will be described below. The acquisition module controls the MICROROC to work in self-trigger mode or external-trigger mode and processes the data from FEB. The trigger module is used in multiple detectors test, where the external trigger signal is necessary. The ADC test module is in charge of the ADC control for the MICROROC debug function described above.

The USB port realized by a USB chip CY7C68013 receives commands from and transmits data to PC. In order to expand in multiple detectors readout system, the SFP port, Ethernet port and DTCC port are reserved as a redundancy design.

\begin{figure}[htb]
\centering
\includegraphics[width=3.5in]{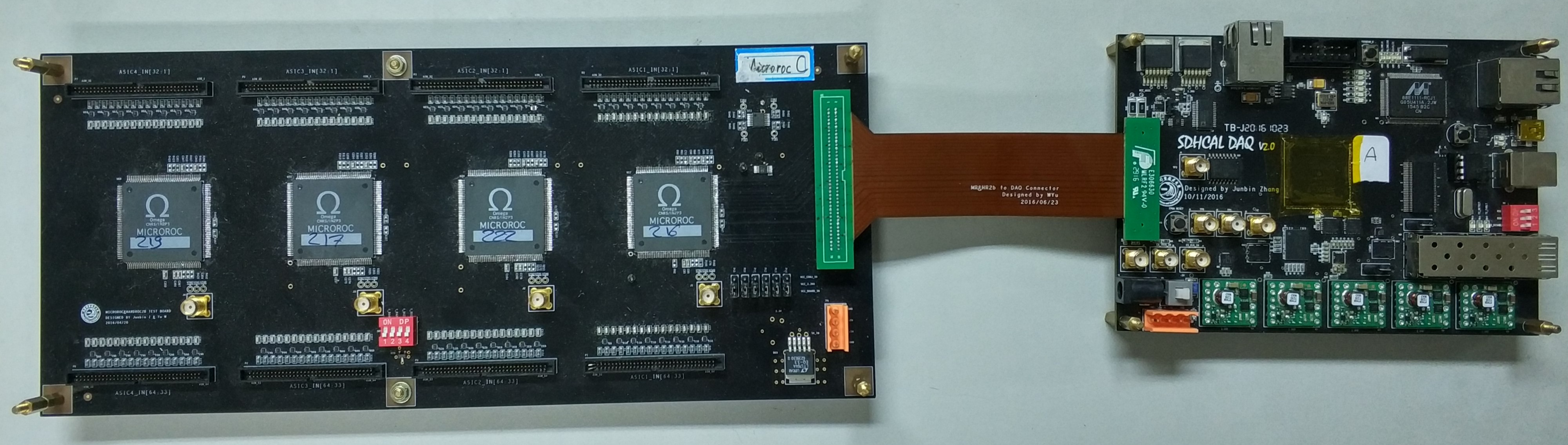}
\caption{The photograph of FEB and DIF.}
\label{fig:Fig5}
\end{figure}

The readout electronics for prototype has been implemented and Fig.~\ref{fig:Fig5} shows the combination of FEB, DIF and their connector plate.

\section{Test and Characterization}
\subsection{Electronics calibration method}
\begin{figure}[htb]
\centering
\includegraphics[width=3.5in]{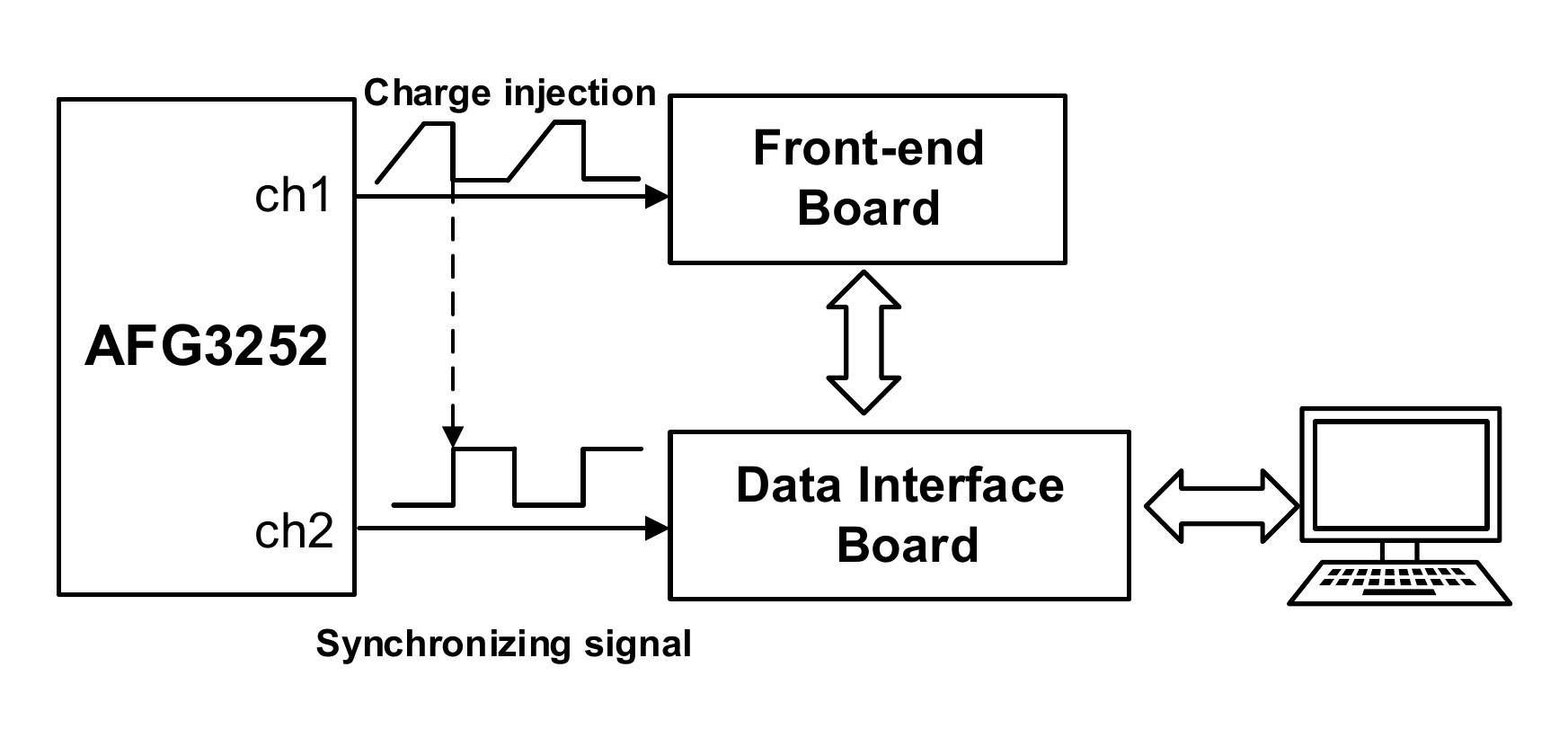}
\caption{Diagram of S-Curve test. Two synchronized signals are sent to FEB and DIF respectively, one for charge injection and another for the count.}
\label{fig:Fig6}
\end{figure}

In order to evaluate the gain and noise of the readout electronics, the digital test signals of MICROROC are used for calibration. The calibration is implemented by injecting known charges to the ASIC and record the state of each discriminator to obtain trigger efficiency. A signal generator (Textronix AFG3252~\cite{bib:bib16}) is employed to inject the charge and synchronous signal (Fig.~\ref{fig:Fig6}). MICROROC can only measure a negative charge, so the charge is injected at the falling edge of the channel 1 through a calibration capacitor and a slow rising edge will prevent the positive charge injection. The value of the charge is calculated as $Q_{in}=\Delta V\times C_{test}$, where $\Delta V$ is the amplitude of channel 1 and  $C_{test}$ is the value of the calibration capacitor. The calibration capacitor is integrated in the MICROROC and the value is 0.5pF. The rising edge of channel 2 is synchronized with the falling edge of channel 1 and this is used to count the injection number. The trigger efficiency is the ratio of discriminator response count to the charge injection count. Changing the threshold of the discriminator with minimum step, the relation between trigger efficiency and threshold can be obtained. This relation can plot an `S' shape curve, so called `S-Curve' method. 

The S-Curve represents the cumulative distribution function of the shaper output signal and the difference of S-Curve is the probability density function of the shaper output. The gain and noise can be deduced from the resulting distribution function. Furthermore, considering the equidistributed noise added on the shaper, the distribution of the shaper output signal obeys a Gaussian distribution. Therefore, the S-Curve can be fitted with the Gauss error function.
\begin{figure}[htb]
\centering
\subfloat[Before alignment, the spread is about 4.5 DAC unit.]{\includegraphics[width=3.5in]{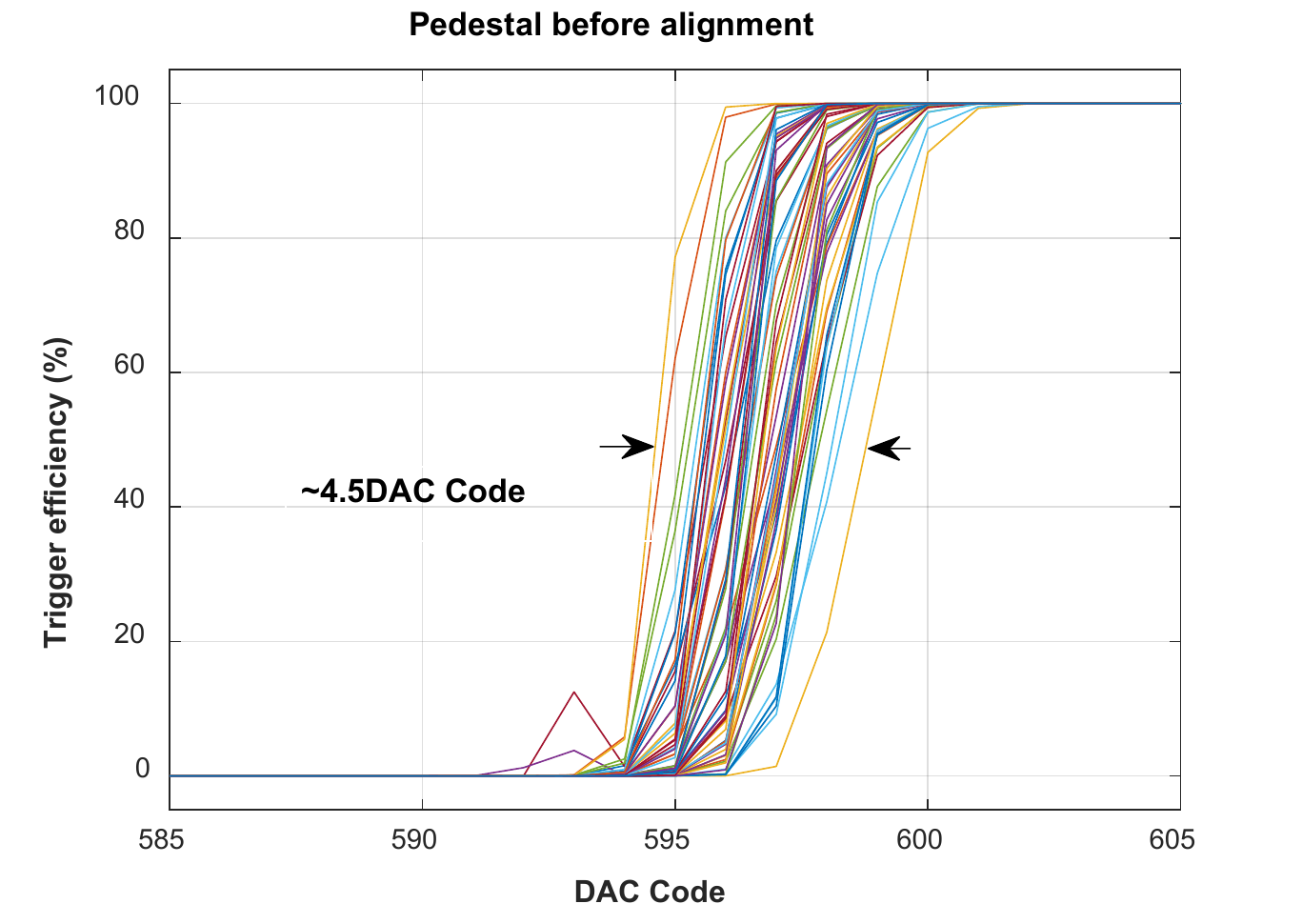}
\label{fig:Fig7a}}
\hspace{0pt}
\subfloat[After alignment, the spread is reduced to 2 DAC unit]{\includegraphics[width=3.5in]{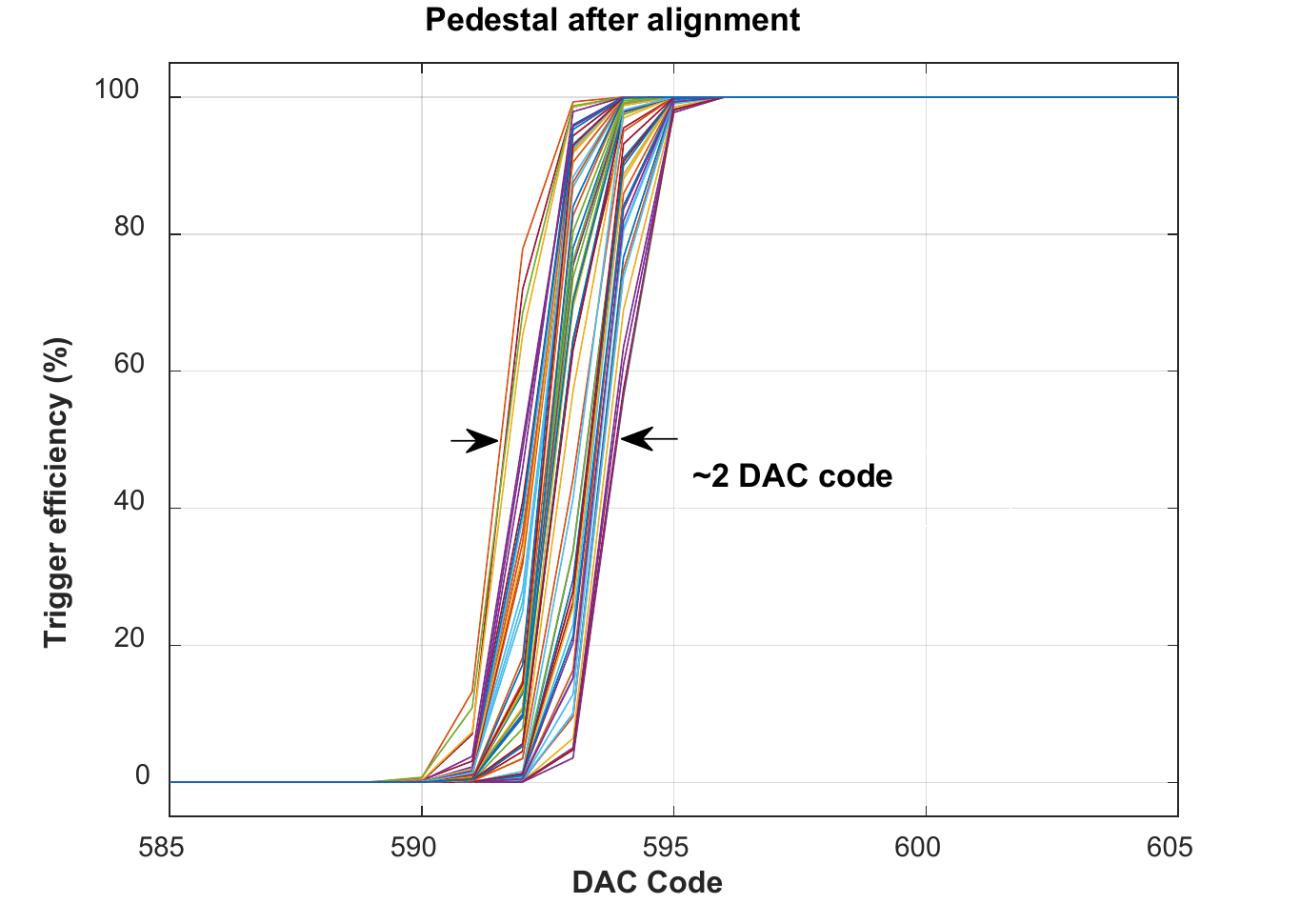}
\label{fig:Fig7b}}
\caption{ S-Curve of the pedestal of one chip before and after alignment.}
\label{fig:Fig7}
\end{figure}
\subsection{Pedestal and gain}
By applying the calibration method, the S-Curve of the pedestal has been obtained (Fig.~\ref{fig:Fig7a}). As the result shows, the pedestal has a spread of about five DAC codes. The threshold DACs are shared by 64 channels. So that, the spread of the pedestal will have a great influence on the lowest threshold of the whole ASIC. To minimize the pedestal spread, each channel of MCIROROC is equipped with an alignment DAC. By setting the align DACs through the slow control parameters, the spread can be reduced into around two DAC codes (Fig.~\ref{fig:Fig7b}).

\begin{figure}[htb]
\centering
\includegraphics[width=3.5in]{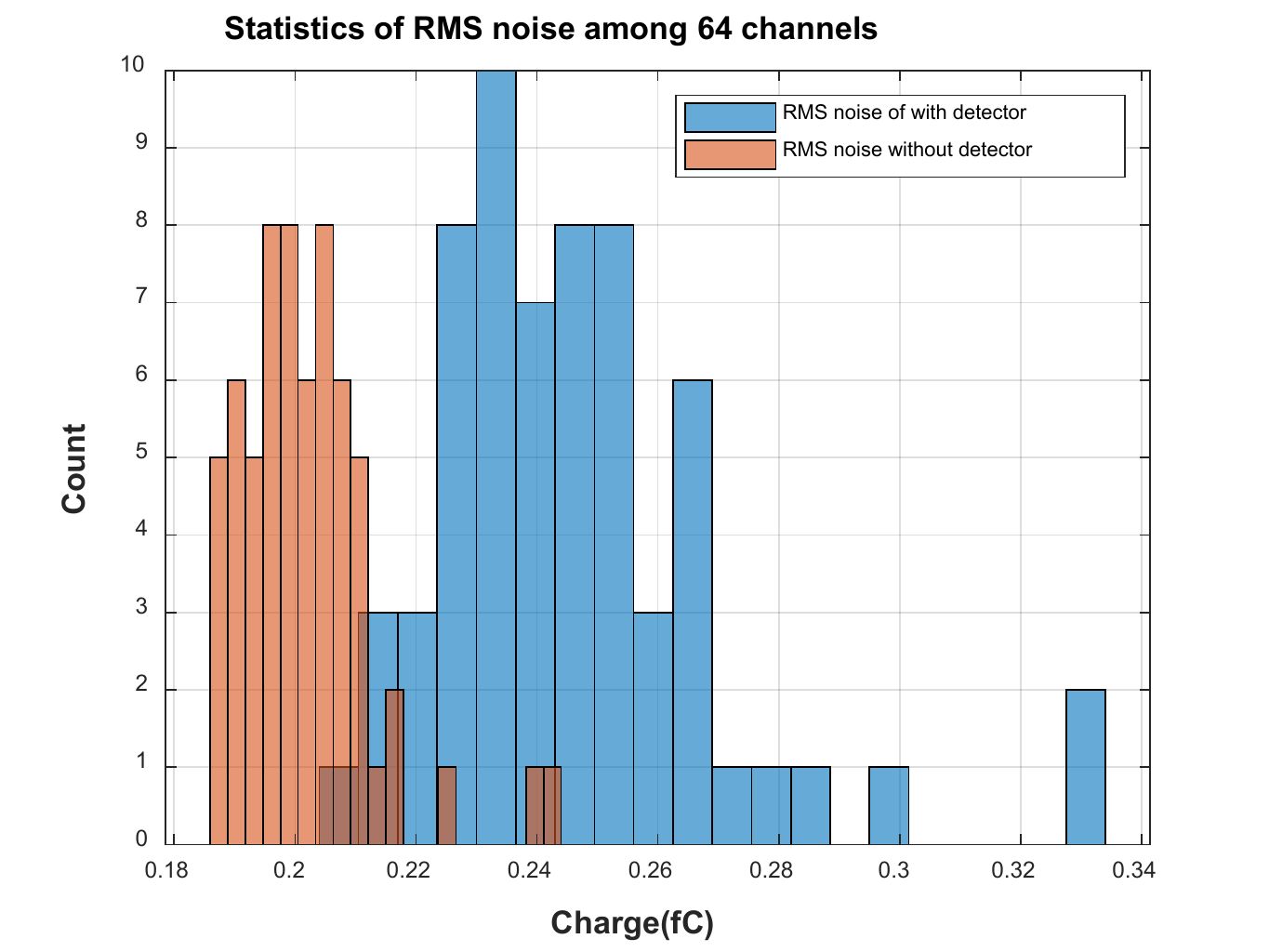}
\caption{RMS noise measurement. The bar in orange is the RMS noise of the electronics with a maximum value of 0.25fC. The blue one is the RMS noise connected to the detector with a maximum value of 0.34fC.}
\label{fig:Fig8}
\end{figure}

After alignment, the noise of the readout electronics has been measured both electronics and with detector (Fig.~\ref{fig:Fig8}). The maximum RMS noise of one channel is 0.34fC, which is much smaller than the minimum ionization particles (MIPs) signal.

\begin{figure}[htb]
\centering
\includegraphics[width=3.5in]{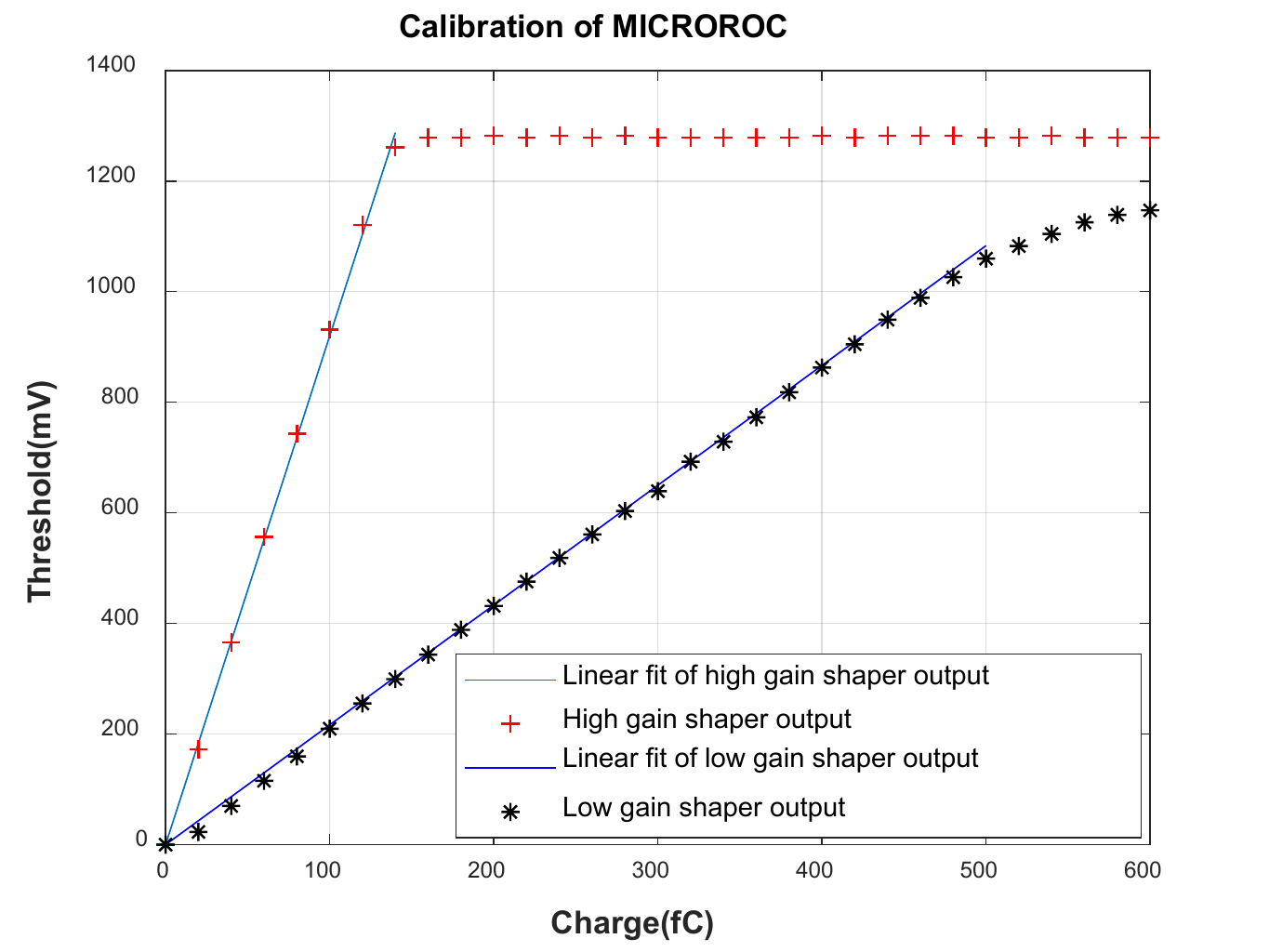}
\caption{Calibration curve with linear fit of the high gain and low gain shaper.}
\label{fig:Fig9}
\end{figure}

The gain is obtained by changing the inject charge from the signal generator. In this procedure, the feedback parameters of two shapers are set as sw\_hg=01 and sw\_lg=01, which can be found in the datasheet \cite{bib:bib17}. The dynamic range of the high gain shaper is 0fC to 140fC with the gain of 9.1961 mV/fC. And the dynamic range of the low gain shaper is up to 500fC with the 2.1663 mV/fC gain (Fig.~\ref{fig:Fig9}). The gain variation of the high gain shaper is below 1\% and this value is corresponding to about 1fC charge within the dynamic range.
\begin{figure}[htb]
\centering
\includegraphics[width=3.5in]{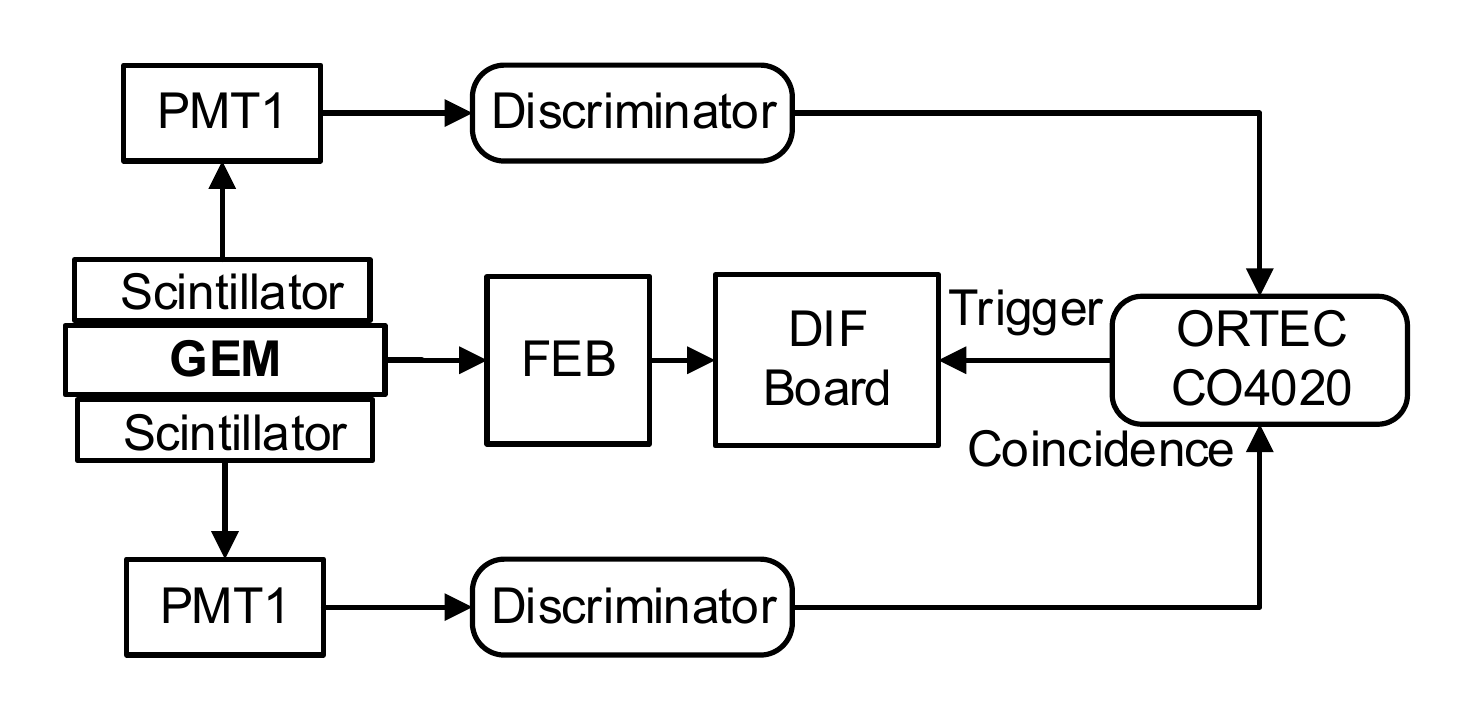}
\caption{Test setup for the cosmic ray test. Two scintillators located above and below the GEM detector are used to generate trigger signal.}
\label{fig:Fig10}
\end{figure}

\subsection{Cosmic-ray test with GEM detector}
\begin{figure}[htb]
\centering
\includegraphics[width=3.5in]{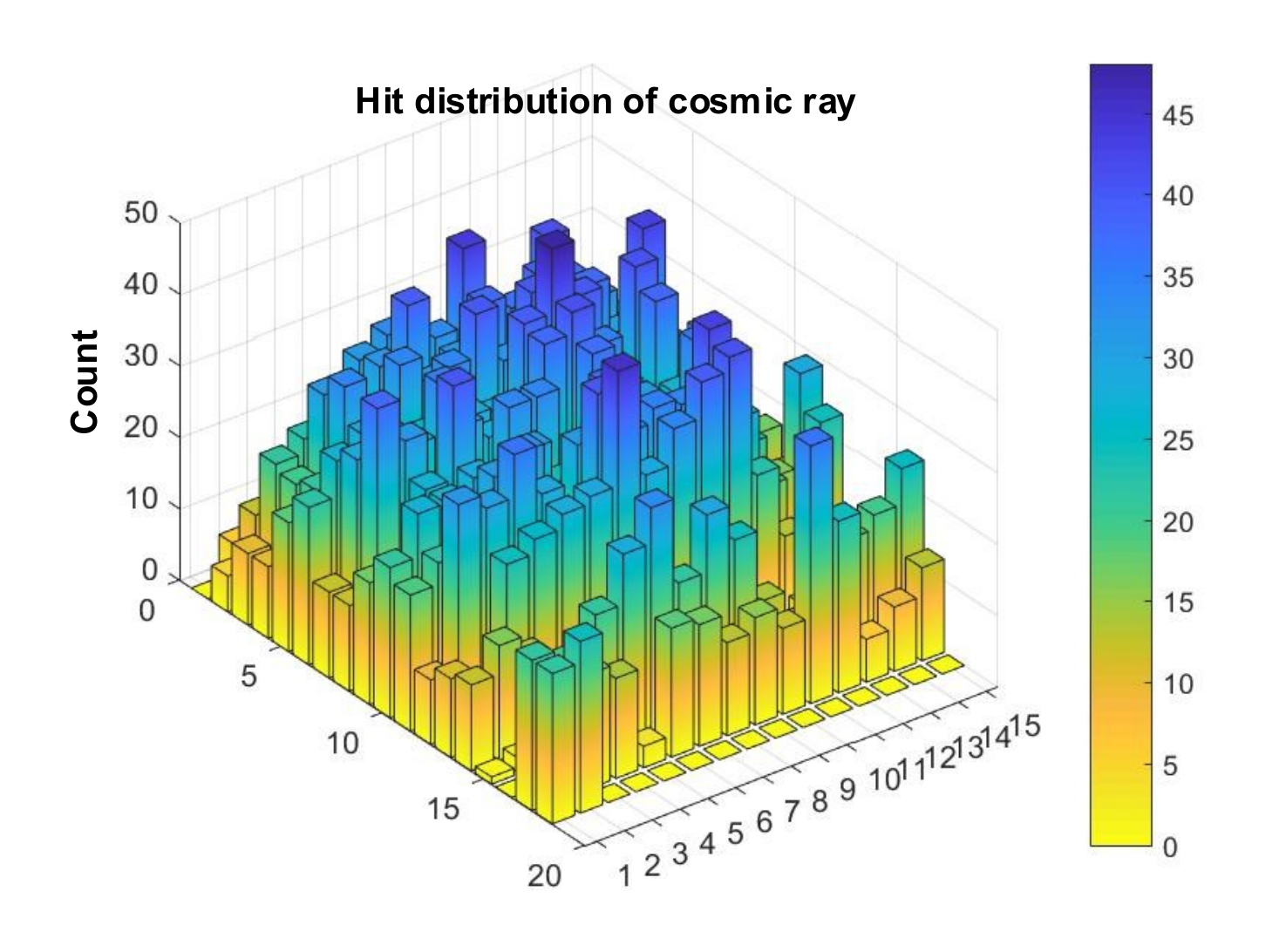}
\caption{Hit distribution of cosmic-ray with coincidence detection. The hit is more concentrated in the coincidence center than the edge.}
\label{fig:Fig11}
\end{figure}

A double-layer GEM detector consisted of 3mm draft gap, 1mm transfer gap and 1mm induction gap is used to demonstrate the performance of the readout electronics. The working gas is 95\%Ar with 5\% iC$_4$H$_{10}$. The test diagram is shown in Fig.~\ref{fig:Fig10}. Two plastic scintillators with PMT readout are used as coincidence detectors and the GEM detector is located between these two plastic scintillators. The signal of each PMT is sent to a discriminator individually and the output signals of two discriminators are connected to a logical unit (ORTEC CO4020~\cite{bib:bib18}). The logical unit is set in coincidence (AND) mode. If a cosmic ray travels through two plastic scintillators, the logical unit will give a trigger signal to the DIF board. Then the DIF board starts the MICROROC and read back hit data. By accumulating the hit data of each pad, the hit distribution is measured (Fig.~\ref{fig:Fig11}). The distribution is more concentrate in the coincidence center, which is in line with the expectations.

For the digital readout calorimeter, one of the factors that influences the energy reconstruction is detection efficiency. The detection efficiency is calculated as the ratio of detector response count to the total particle count~\cite{bib:bib19}. The efficiency is mainly influenced by the detector geometry, gain of the detector and the electronics threshold. A higher efficiency means that less particles are missing.

\begin{figure}[htb]
\centering
\includegraphics[width=3.5in]{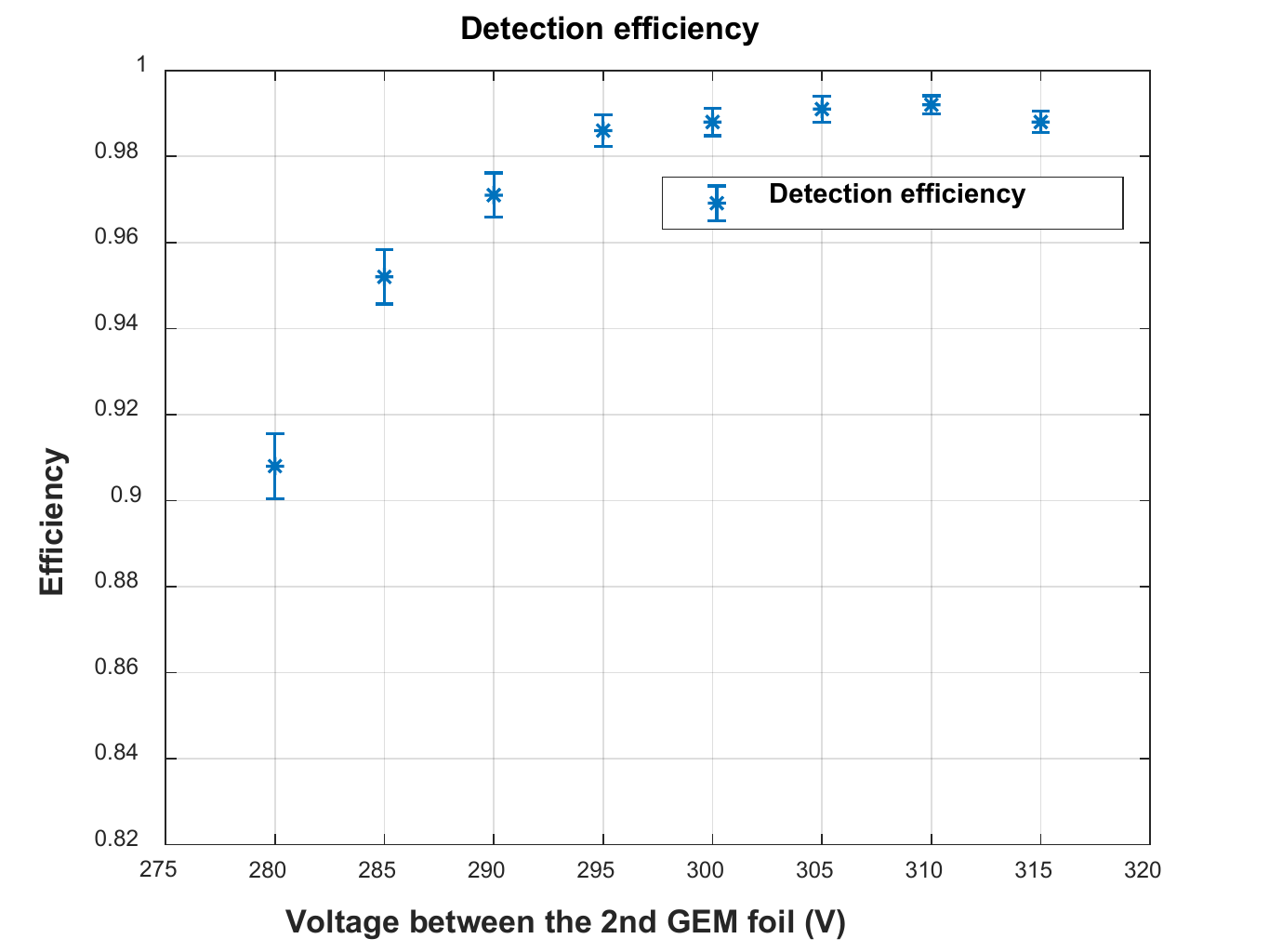}
\caption{Detection efficiency of GEM with different gain. As the gain grows, the efficiency reaches to the plateau region.}
\label{fig:Fig12}
\end{figure}

As shown in Fig.~\ref{fig:Fig10}, the DIF board can both record the valid hit data and total count of particles. The electronics threshold is 5fC to reject the noise and the gain of the detector is adjusted by changing the voltage between the GEM foil. The detection efficiency with different gas gain is shown in Fig.~\ref{fig:Fig12}. With the gain increasing, the efficiency reaches to the plateau region with the value about 98\%. The high efficiency meets the requirement of the calorimeter and indicates that the electronics works well with detector.

\subsection{X-ray test}
Gain of the detector is tested with the 8keV X-ray with the external ADC module. This module utilizes the test function and trigger output function of MICROROC. When a trigger signal from MICROROC is detected, this module sends a hold signal to maintain the peak of the low gain shaper output signal. Once the hold signal is received, an external ADC starts sampling output signal of the shaper.

\begin{figure}[htb]
\centering
\includegraphics[width=3.5in]{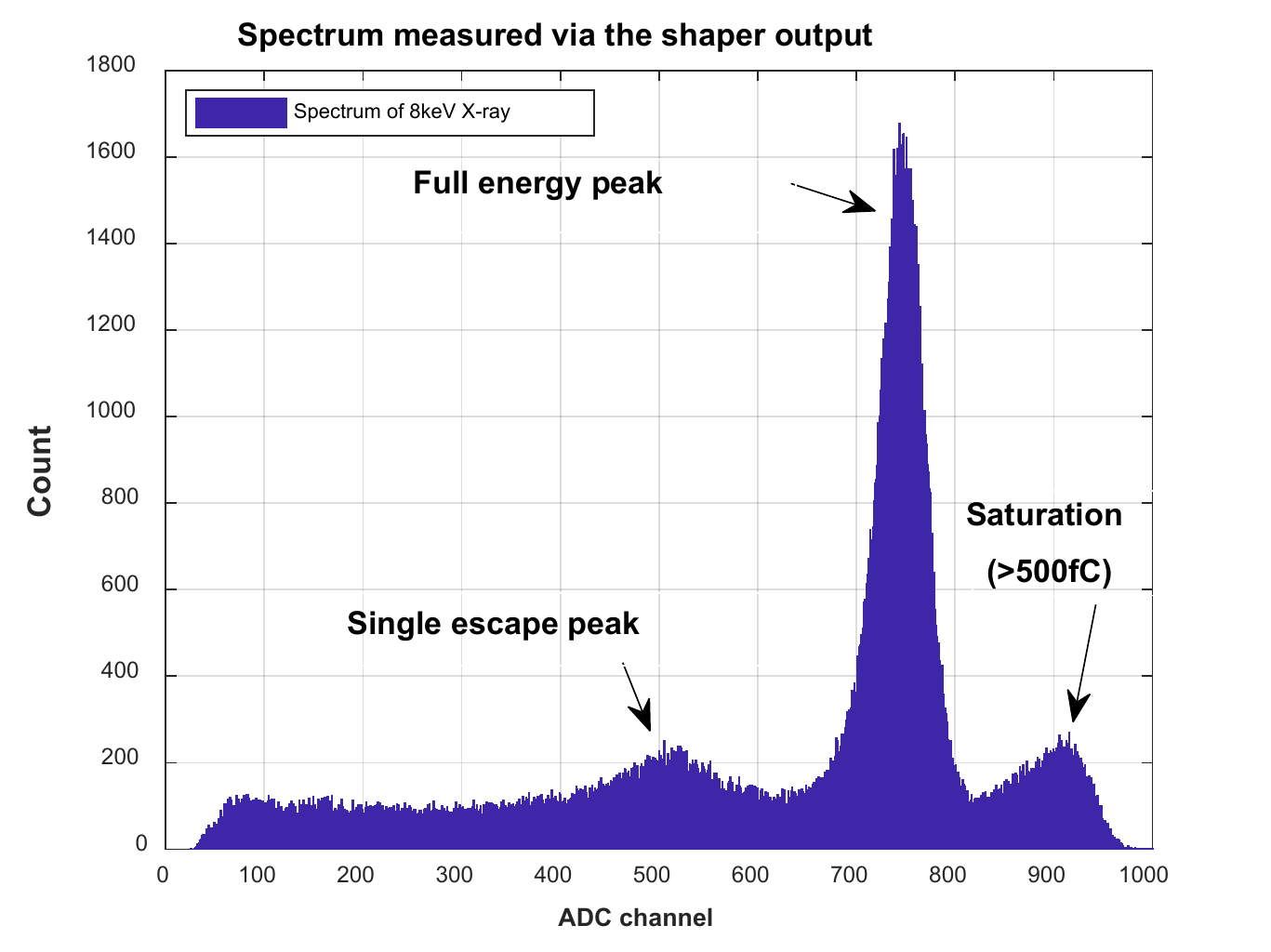}
\caption{Spectrum of 8keV X-ray measured by external ADC.}
\label{fig:Fig13}
\end{figure}

The energy spectrum of 8keV X-ray is measured by setting a proper threshold to cut off the noise (Fig.~\ref{fig:Fig13}). Due to some large signal over the dynamic range of the MICROROC, a saturation is observed. However, this saturation does not influence the full energy peak.

By measuring the full energy peak of the readout pads, the gain of the detector can be evaluated. As measured in the former test \cite{bib:bib20}, the gain would not change rapidly in a small area. So that we measure the gain of the detector every second pad to reduce the test time. Test results show that the uniformity is better than 20\% which is consistent with the former result \cite{bib:bib20}.

\section{Conclusion}
This paper presents a semi-digital readout electronics for the GEM detector, which is one of the candidates for the semi-digital hadron calorimeter. A front-end board with 256 readout channels and a data interface board with multiple readout port are designed and implemented. Test results show that the noise is much lower than the MIP signal and the dynamic range can cover a large energy range, which is enough for the hadron measurement.

Eventually, this readout electronics is applied on a double-layer GEM detector. The cosmic ray events can be successfully recorded without false trigger and the readout electronics cooperates with the GEM detector well.

\section*{Acknowledgment}
Thanks to St\'ephane CALLIER and Christophe de LA TAILLE at OMEGA - IN2P3/CNRS for helpful discussions and useful suggestions.

\ifCLASSOPTIONcaptionsoff
  \newpage
\fi



%\bibliographystyle{IEEEtran}


\end{document}